\newif{\ifarxiv}
\newif{\ifdraft}
\newif{\ifremarks}

\arxivtrue  

\documentclass[11pt,a4paper]{article}
\usepackage{lmodern}
\usepackage[T1]{fontenc}
\usepackage{microtype}
\usepackage[left=2.8cm,right=2.8cm,top=2.8cm,bottom=2.8cm]{geometry}
\usepackage{graphicx}
\usepackage{amsmath,amssymb,mathtools}
\usepackage[bookmarks=true]{hyperref}
\usepackage{cite}
\usepackage[bulletsep]{collref} 
\usepackage{xcolor}

\ifremarks
\newcommand{\remarkt}[1]{{\renewcommand{\bfdefault}{b}{\color[RGB]{0,150,0}{\textbf{T: #1}}}}}
\newcommand{\remarkg}[1]{{\renewcommand{\bfdefault}{b}{\color[RGB]{200,0,100}{\textbf{G: #1}}}}}
\newcommand{\remarkv}[1]{{\renewcommand{\bfdefault}{b}{\color[RGB]{0,70,170}{\textbf{V: #1}}}}}
\fi
\providecommand{\remarkt}[1]{\ignorespaces}
\providecommand{\remarkg}[1]{\ignorespaces}
\providecommand{\remarkv}[1]{\ignorespaces}

\usepackage[font=small,labelfont=bf,width=0.85\textwidth]{caption}

\providecommand{\hypersetup}[1]{}
\providecommand{\hyperref}[2][]{#2}
\providecommand{\texorpdfstring}[2]{#1}
\providecommand{\pdfbookmark}[3][]{}

\hypersetup{plainpages=false}
\hypersetup{pdfpagemode=UseOutlines}
\hypersetup{bookmarksnumbered=true}
\hypersetup{bookmarksopen=true}
\hypersetup{pdfstartview=FitH}
\hypersetup{colorlinks=false}
\hypersetup{citebordercolor={.6 .9 .6}}
\hypersetup{urlbordercolor={.7 .8 1}}
\hypersetup{linkbordercolor={1 .7 .7}}
\hypersetup{pdfborder={0 0 .5}}

\newcommand{\namedref}[2]{\hyperref[#2]{#1~\ref*{#2}}}
\newcommand{\secref}[1]{\namedref{Section}{#1}}
\newcommand{\appref}[1]{\namedref{Appendix}{#1}}

\newcommand{\figref}[1]{\namedref{Figure}{#1}}

\makeatletter
\def\mr@ignsp#1 {\ifx\:#1\@empty\else #1\expandafter\mr@ignsp\fi}%
\newcommand{\multiref}[1]{\begingroup
\xdef\mr@no@sparg{\expandafter\mr@ignsp#1 \: }%
\def\mr@comma{}%
\@for\mr@refs:=\mr@no@sparg\do{\mr@comma\def\mr@comma{,}\ref{\mr@refs}}%
\endgroup}
\makeatother
\renewcommand{\eqref}[1]{(\multiref{#1})}

\makeatletter
\let\@myabstract\@empty
\let\@keywords\@empty
\let\@subject\@empty
\providecommand{\affiliation}[1]{\gdef\@affiliation{#1}}
\providecommand{\myabstract}[1]{\gdef\@myabstract{#1}}
\providecommand{\keywords}[1]{\gdef\@keywords{#1}}
\providecommand{\subject}[1]{\gdef\@subject{#1}}
\def\thetitle{\@title}
\def\theauthor{\@author}
\def\theaffiliation{\@affiliation}
\def\theabstract{\@myabstract}
\def\thesubject{\@subject}
\def\thedate{\@date}
\def\thekeywords{\@keywords}
\makeatother
\def\fillpdfdata{
\hypersetup{pdftitle={\thetitle}}%
\hypersetup{pdfsubject={\thesubject}}%
\hypersetup{pdfkeywords={\thekeywords}}%
}

\let\oldbfseries=\bfseries
\let\oldmdseries=\mdseries
\let\oldnormalfont=\normalfont
\renewcommand{\bfseries}{\oldbfseries\boldmath}
\renewcommand{\mdseries}{\oldmdseries\unboldmath}
\renewcommand{\normalfont}{\oldnormalfont\unboldmath}

\numberwithin{equation}{section}

\makeatletter
\newlength{\apb@width}
\newcommand{\autoparbox}[2][c]{\settowidth{\apb@width}{#2}\parbox[#1]{\apb@width}{#2}}

\makeatother

\newcommand{\nn}{\nonumber}

\newcommand{\brk}[1]{(#1)}
\newcommand{\lrbrk}[1]{\left(#1\right)}
\newcommand{\bigbrk}[1]{\bigl(#1\bigr)}

\newcommand{\sbrk}[1]{[#1]}

\newcommand{\brc}[1]{\{#1\}}

\newcommand{\smb}[1]{(#1)}

\newcommand{\biggsmb}[1]{\biggl(#1\biggr)}
\newcommand{\br}[1]{\langle#1\rangle}

\newcommand{\supup}[1]{^{\mathrm{#1}}}

\newcommand{\superN}{\mathcal{N}}
\newcommand{\grp}[1]{\mathrm{#1}}

\newcommand{\diag}{\operatorname{diag}}
\newcommand{\disc}{\operatorname{disc}}
\newcommand{\dlog}{\operatorname{dlog}}

\newcommand{\smbop}[1]{\operatorname{S}\sbrk{#1}}
\newcommand{\Reals}{\mathbb{R}}
\newcommand{\order}[1]{\mathcal{O}(#1)}

\newcommand{\mathematica}{\textsc{Mathematica}}

\begin{document}

\ifdraft
\addtolength{\baselineskip}{3pt}
\fi


\title{The Two-Loop Symbol of all Multi-Regge Regions}

\myabstract{We study the symbol of the
two-loop $n$-gluon MHV amplitude for all Mandelstam regions in
multi-Regge kinematics in $\superN=4$ super Yang--Mills theory.
While the number of distinct Mandelstam regions grows exponentially
with $n$, the increase of independent symbols turns out to be merely
quadratic. We uncover how to construct the symbols for any number of
external gluons from just two building blocks which are naturally
associated with the six- and seven-gluon amplitude, respectively. The
second building block is entirely new, and in addition to its symbol,
we also construct a prototype function that correctly reproduces all
terms of maximal functional transcendentality.}



\ifarxiv
\noindent
\mbox{}\hfill DESY 15-257\\
\mbox{}\hfill SLAC-PUB-16473
\fi

\author{%
	Till Bargheer,\texorpdfstring{${}^a$}{}
	Georgios Papathanasiou,\texorpdfstring{${}^b$}{}
	and Volker Schomerus\texorpdfstring{${}^a$}{}%
}

\hypersetup{pdfauthor={\theauthor}}

\vfill

\begin{center}
{\Large\textbf{\mathversion{bold}\thetitle}\par}
\vspace{1cm}

\textsc{\theauthor}
\vspace{5mm}

\textit{%
${}^a$DESY Theory Group, DESY Hamburg, Notkestra{\ss}e 85, D-22607 Hamburg, Germany\\
${}^b$SLAC National Accelerator Laboratory, Stanford University, Stanford, CA 94309, USA}

\vspace{3mm}
{\ttfamily
\href{mailto:till.bargheer@desy.de}{till.bargheer@desy.de},
\href{mailto:georgios@slac.stanford.edu}{georgios@slac.stanford.edu},
\href{mailto:volker.schomerus@desy.de}{volker.schomerus@desy.de}}
\par\vspace{1cm}

\textbf{Abstract}\vspace{5mm}

\begin{minipage}{12.7cm}
\theabstract
\end{minipage}

\vspace{1cm}

\end{center}

\fillpdfdata

\hrule
\providecommand{\microtypesetup}[1]{}
\microtypesetup{protrusion=false}
\tableofcontents
\microtypesetup{protrusion=true}

\vspace{3ex}\hrule

\vfill

\newpage

\section{Introduction}

Constructing the S-matrix of a 4D gauge theory, even in the planar limit,
is one of the central challenges of theoretical physics. In the case of
$\superN=4$ supersymmetric Yang--Mills (SYM) theory, dual conformal
symmetry~\cite{Drummond:2006rz,Drummond:2007au} uniquely fixes the
form of scattering amplitudes of up to $5$ external gluons to coincide
with the  BDS ansatz~\cite{Bern:2005iz}. But it allows for an
additional finite part that can depend on $3n-15$ conformal cross
ratios for amplitudes with $n\ge 6$ gluons. Impressive progress towards
the determination of the latter quantity has been made, in particular
for $n=6$ external gluons. In this case, there exists an all-loop formula
which determines the remainder function for general
kinematics~\cite{Basso:2015uxa}. A number of complementary approaches
paved the way for this important result. These included perturbative
studies in special kinematic regions, such as the collinear and the
Regge limit, see~\cite{Alday:2010ku,Gaiotto:2010fk,Gaiotto:2011dt}
and~\cite{Lipatov:2010ad,Bartels:2011ge,Bartels:2011xy} for early
contributions. Their  findings provided essential boundary conditions
into the amplitude bootstrap for fixed-order calculations in general
kinematics that was initiated in~\cite{Dixon:2011pw}
(see also~\cite{Dixon:2014xca} for a recent overview and more
references). In a series of
papers~\cite{Basso:2013vsa,Basso:2013aha,Basso:2014jfa,Basso:2014pla,Basso:2015rta,Basso:2015uxa},
Basso et al.\ then developed the non-perturbative Wilson loop OPE,
and showed that it could accommodate all previous results on hexagon amplitudes
and even correctly interpolate to strong coupling where string theory
takes over~\cite{Alday:2007hr,Alday:2009dv,Alday:2010vh,Bartels:2010ej,Bartels:2013dja}.

While some partial results are known in particular for $n=7$, see
e.g.~\cite{Bartels:2013jna,Golden:2014xqf,Drummond:2014ffa,Bartels:2014jya,Bartels:2014ppa}
and references therein, the scattering problem of $\superN=4$ SYM
theory with more than six gluons has not been solved. In pushing the
entire program to higher
numbers of external gluons and uncovering universal patterns, the multi-Regge
limit of high-energy gluon scattering is expected to play an important
role. Similar to the Wilson loop OPE, the expansion around multi-Regge
kinematics is based on elementary building blocks, such as BFKL eigenvalues,
impact factors and production vertices. These are subject to powerful
constraints from integrability which have been partially worked out, see
e.g.~\cite{Bartels:2011nz}. In addition, the multi-Regge limit was
shown~\cite{Bartels:2012gq} to correspond to the infrared limit of the
auxiliary
one-dimensional integrable system that controls the strongly coupled
theory~\cite{Alday:2010vh}. In this regime, the quantum fluctuations
of the auxiliary
system are suppressed, which turns the original system of coupled non-linear
integral equations into much simpler algebraic equations for Bethe roots.

Turning to the amplitude bootstrap, we recall that it relies on the
observation that $L$-loop amplitudes of certain helicity are expressed
in terms of multiple polylogarithm functions~\cite{Goncharov:2001iea} of
transcendental weight $2L$ (see~\cite{Duhr:2014woa} for a review). This is
backed by the ``dlog'' representation of the all-loop
integrand~\cite{ArkaniHamed:2012nw}, as well as the perturbative
analysis of the OPE
expansion~\cite{Papathanasiou:2013uoa,Papathanasiou:2014yva,Drummond:2015jea}.
However, except for six and seven particles~\cite{Golden:2013xva}, to
date we know of no principle that would motivate from which set of
``letters'' (or
``alphabet'') these multiple polylogarithms can draw their arguments.
The multi-Regge limit could serve as a stepping stone in this direction,
since the kinematical dependence simplifies considerably, and the
experience from the six-gluon analysis suggests that amplitudes
inherit special analytic properties in this limit~\cite{Dixon:2012yy}.
These properties are also expected to make the evaluation of the
integral formulas describing the amplitude in the limit much
simpler than the ones arising in the OPE approach around collinear kinematics, where the
question of resummation at weak and strong coupling represents a
formidable task~\cite{Papathanasiou:2014yva,Basso:2014jfa,Bonini:2015lfr,Belitsky:2015lzw}.

With these goals in mind, we will focus on the multi-Regge limit of the
simplest, Maximally Helicity Violating (MHV) $n$-gluon amplitudes with
all but two of the helicities being the same. For these amplitudes, the
finite part not fixed by dual conformal symmetry is a single
\emph{remainder function} $R_n$. Although the multi-Regge limit of the latter
vanishes in the Euclidean region where all Mandelstam invariants are
spacelike~\cite{Bartels:2008ce,DelDuca:2008jg}, it possesses a rich
set of branch cuts. Exploring the
branch structure through analytic continuation in the Mandelstam
invariants leads to various Mandelstam regions with non-trivial multi-Regge limit.
The number of different regions increases exponentially with the
particle number, hence it is natural to ask for the simplest subset of
regions that contains all the independent
``boundary data'' to be used for constructing the amplitude
in general kinematics.

In this note we consider the $2\rightarrow (n-2)$ multi-Regge limit in
$2^{n-4}$ different Mandelstam regions, which are reached by analytic
continuation in the momenta of any combination of $(n-4)$ adjacent external
particles from positive energy (in the Euclidean region) to negative energy.
Our starting point is the known two-loop $n$-point MHV symbol~\cite{CaronHuot:2011ky}.
Symbols~\cite{Goncharov.A.B.:2009tja,Goncharov:2010jf,Duhr:2011zq,Duhr:2014woa}
capture the most complicated part of the amplitude with the highest
\emph{functional} transcendental weight in a way that trivializes all
identities among (multiple) polylogarithms. The investigation of the
multi-Regge limit of two-loop symbol was initiated in~\cite{Prygarin:2011gd}, restricted to
the leading term in the multi-Regge limit (leading logarithmic approximation, or
LLA), and for a single Mandelstam region. Our analysis extends both
aspects: We consider all $2^{n-4}$ Mandelstam regions, and we include
the first subleading term in the multi-Regge limit (NLLA). As
results, we shall find that \textit{(i)} all independent information is
contained in a subset of only $(n-4)(n-5)/2$ regions and \textit{(ii)}
in all regions the multi-Regge limit of the symbol decomposes into two
basic building blocks $f$ and $g$, naturally associated to the $n=6$
and $n=7$ symbols, respectively. The first building block $f$ was
already discussed in~\cite{Bartels:2011ge,Prygarin:2011gd} to LLA, and
here we identify it to NLLA. The second object $g$, which receives
contributions only from NLLA, is entirely new. Apart from spelling it
out explicitly, we also find a functional representative for it,
belonging to the class of two-dimensional harmonic polylogarithms
(2dHPLs)~\cite{Gehrmann:2000zt}. We complete
the construction through a prescription for how to build the two-loop symbol in the
multi-Regge limit of $2^{n-4}$ Mandelstam regions for any number of external
gluons from the building blocks $f$ and $g$.

This article is organized as follows. In \secref{sec:symbols} we
briefly review some basic facts about iterated integrals, symbols and
their discontinuities. We apply these in \secref{sec:regions} to
obtain the form of the two-loop $n$-point symbol in the different
Mandelstam regions, after also discussing how these can be reached by
analytically continuing the kinematical invariants. In
\secref{sec:mrlrelations} we take the multi-Regge limit of the symbol,
and show how the answer in any region may be reconstructed from the
simplest regions in
which only a single branch cut contributes to the multi-Regge limit.
One of the main results of the paper, the decomposition of the symbol
into the two building blocks for these regions, is the subject of
\secref{sec:blocks}. \secref{sec:function} deals with the uplift of
the newly found seven-gluon building block to a function, whereby we
uniquely fix the maximal transcendental part, and further constrain
the possible terms of lower transcendentality by symmetry.
\secref{sec:conclusions} contains our conclusions. In
\appref{app:mrlparam}, we present a particular parametrization of the
kinematics in terms of momentum twistors that we found very useful for
our analysis.

Computer-readable files with our results accompany
our article on the \texttt{arXiv}. This data constitutes an important
step towards determining the new quantity appearing in the
Balitsky-Fadin-Kuraev-Lipatov (BFKL) approach for $R_7$ known as the
central emission vertex, that so far has only been computed to
LLA~\cite{Bartels:2013jna,Bartels:2014jya}. Furthermore, the
decomposition we have discovered is suggestive of a factorization
structure that may impose new constraints on the analogous BFKL
quantities also appearing at higher points.

\section{Symbols and Discontinuities}
\label{sec:symbols}
\def\x{\text{x}}
\def\disc{\text{Disc}}

Let us consider the $(3n-15)$-dimensional space $\cal X$ of independent dual conformal
invariant cross ratios for an $n$-gluon scattering process, and a curve $\gamma:
[0,1]\rightarrow {\cal X}$ in the space of kinematic invariants.
It starts at the base point $\gamma(0) \in {\cal X}$ and can run to any
point $\x = \gamma(1) \in {\cal X}$. At two loops, the remainder
function is a sum of iterated integrals~\cite{Chen:1977oja} of the
form
\begin{equation}
R(\x) \sim
\int\limits_{\mathclap{0\leq t_1\leq\dots\leq t_4\leq 1}}
\dlog(X_{a_1}(t_1))\dots\dlog(X_{a_4}(t_4))\,,
\label{eq:iteratedint}
\end{equation}
where $X_a$ is some set of functions on ${\cal X}$, indexed by
some finite index set, i.e.\ $a = 1,\dots,\varpi$, which depend on the
parameter $t \in [0,1]$ through the curve $\gamma$. In order
to keep notations simple we have only displayed a single summand. The symbol
$\sim$ should remind us of the fact that the true remainder function $R$ is
composed from a finite sum of such integral contributions. All of them
contain four integrations, i.e.\ they possess the same transcendentality
degree, or weight.

In order for the integral to be well-defined, we should avoid curves
$\gamma$ which pass through the set ${\cal Z}$ of zeros and singularities
of $X_a$. Consequently, $R(\x)$ is only defined for $\x \in {\cal Y} \equiv
{\cal X}\setminus {\cal Z}$. There exists an integrability condition, which
insures that an iterated integral depends only on the base point $\gamma(0)$
and the homotopy class $[\gamma]$ of $\gamma$ in ${\cal Y}$~\cite{Chen:1977oja}.
Provided the integrability condition holds, and given a base point $\gamma(0)$, the
integral $R$ defines a multivalued function on ${\cal Y}$. It has branch
cuts $B_\nu$ which end at the zeros and singularities of the entries $X_a$. The
discontinuities $\disc_\nu$ along the branch cuts are given by
\begin{multline}
\disc_{\nu}(R(\x))\sim
\\
{\sum}_{j=1}^{4}
\ \int\limits_{\mathclap{0\leq s_{1}\leq\dots\leq s_j\leq 1}}
\dlog(X_{a_1}(s_{1}))\dots\dlog(X_{a_j}(s_j))
\,
\int\limits_{\mathclap{0\leq t_{j+1}\leq\dots\leq t_4\leq 1}}
\dlog(X_{a_{j+1}}(t_{j+1}))\dots\dlog(X_{a_4}(t_4))
\,.
\end{multline}
In this expression, the parameters $t_i$ are mapped into the space ${\cal Y}$
of kinematic invariants through the curve $\gamma$, as before, while $s_i$
are sent to ${\cal Y}$ through a closed curve $\eta_\nu=\eta_\nu(s)$ which
starts and ends at $\gamma(0)$, winds around the branch point $B_\nu$ once,
and avoids winding around all other $B_\mu$, $\nu \neq \mu$. Consequently, the
integral on the left in the above equation evaluates to a number,  i.e.\ it does not depend on the
endpoint $\x$ of the curve $\gamma$. All dependence on $\x$ comes in through
the iterated integrals on the right. In order to derive this expression,
one moves the base point $\gamma(0)$ along the closed curve $\eta_\nu$. The
iterated integral along the concatenation $\gamma\circ \eta$ decomposes into
a sum of products of iterated integrals of lower functional transcendentality.
By definition, the discontinuity is the difference between the integral for
$\gamma \circ \eta$ and the original $\gamma$. It is a linear combinations of
iterated integrals of functional transcendentality degree less or equal to three.

The symbol $\smbop{R}$ is a linear map on the space of iterated integrals
which is defined such that~\cite{Goncharov.A.B.:2009tja}
\begin{equation}\label{eq:symbol}
S[R(\x)] \sim
\smb{X_{a_1}\otimes\dots\otimes X_{a_4}}\ .
\end{equation}
Note that the symbol forgets all information encoded in the choice
of the base point and path. Hence, it determines the original iterated
integral $R$ only up to certain functions of lower transcendentality. On the
other hand, it knows about the endpoints of branch cuts and allows to determine
the symbol of the corresponding discontinuities,
\begin{equation}
S[\disc_\nu(R(\x))] \sim \left[
\oint_{\eta_\nu} \dlog(X_{a_1})\right]\  \lrbrk{X_{a_2} \otimes \cdots \otimes X_{a_4}}
 \,,
\end{equation}
From the right hand side, we can reconstruct the discontinuity up to
certain functions of transcendentality degree less or equal to two.

Before we conclude this section, let us make one more comment. The
discontinuities across branch cuts provide a representation of the
homotopy group $\pi_1({\cal Y})$. On the other hand, commutators of
elements in the homotopy group are related to double discontinuities,
and hence these are expressed through iterated integrals of
transcendentality at most two. Such integrals do not show up in a
symbol of length three and hence we will be able to safely ignore
the difference between homotopy and homology in the following
analysis.

\section{Mandelstam regions and cuts}
\label{sec:regions}

A relatively simple expression for the two-loop remainder functions has
been found for $n=6$ in~\cite{Goncharov:2010jf}. The two-loop remainder
function was also determined for $n=7$ in~\cite{Golden:2014xqf}, but the
expressions are quite complex already. On the other hand, the symbol of
the two-loop remainder function is actually known for any number $n$ of
gluons~\cite{CaronHuot:2011ky}. As we have just recalled, this symbol
determines the finite remainder $R^{(2)}_n$ up to functions of
transcendentality at most three. Our goal is to analyze this
symbol, and in particular the symbol of its discontinuities,
in the multi-Regge limit.

\begin{figure}[htb]
\centering
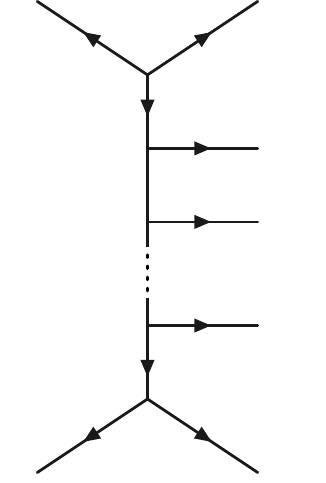
\qquad\qquad
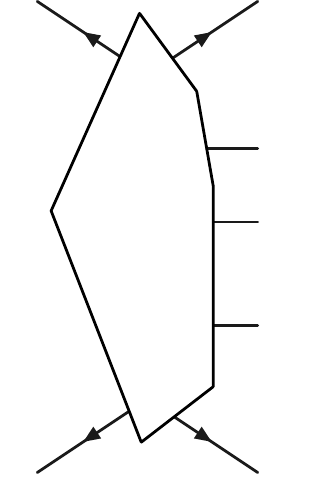
\caption{Kinematics of the scattering process $2 \to n-2$. On the
right-hand side we show a graphical representation of the dual
variables $x_i$.}
\label{fig:configuration}
\end{figure}

As we reviewed above, to leading transcendental order the remainder
function has branch cuts that end at the zeros and infinities of those
functions $X_a$ that appear as a first entry of the symbol. The
positions of these branch cuts are dictated by unitarity to coincide
with thresholds where an intermediate particle goes on shell. For
planar massless theories, this can only happen when a sum of
cyclically adjacent external momenta becomes
null~\cite{Gaiotto:2011dt}. Since dual conformal invariance
additionally constrains the remainder function to only depend on
conformal invariant combinations of Mandelstam invariants, from these
considerations we deduce that the first entries of the symbol can only
be cross ratios of the form
\begin{equation}\label{eq:Uij}
 U_{ij}=\frac{x^2_{i+1,j}x^2_{i,j+1}}{x^2_{ij}x^2_{i+1,j+1}} \ .
\end{equation}
These are defined for $i,j= 1, 2, \dots n$ and $3 \leq |i-j| \leq
n-2$ in terms of the distances
\begin{equation}
x_{ij}^2=(x_i-x_j)^2
\end{equation}
between the cusps $x_i$ of the usual light-like polygon that encodes
the kinematics of the scattering event through $p_i = x_i-x_{i-1}$. Our
conventions concerning the enumeration of gluons are shown in
\figref{fig:configuration}. The black dots on the right hand side depict
the cusps. Counting the different possibilities in eq.~\eqref{eq:Uij} we
see that we can have a total of $n(n-5)/2$ distinct cross ratios.
While $(n-4)(n-5)/2$ of them are given by the cross ratios
\begin{equation}
u_{ij} = U_{ij}
\quad\text{with}\quad
i = 2,\dots, n-4,
\quad
j = i+3, \dots,n-1\,,
\end{equation}
it turns out convenient to use $2(n-5)$ products and quotients of
the remaining cross ratios $U_{1j}$ and $U_{j-1,n}$,
\begin{equation}\label{eq:ut_eps_crossratios}
\tilde u_j := U_{1j} U^{-1}_{j-1,n}\,,
\qquad
\varepsilon_j := U_{1j} U_{j-1,n}
\end{equation}
where $j=4,\dots, n-2$. Of course, the usual rules of the symbol
calculus allow to pass easily from $\tilde u_j$ and $\varepsilon_j$
to the more conventional first entries $U_{1j}$ and $U_{j-1,n}$.

Our goal is to analyze the discontinuities along the cuts that end at
points where one of the kinematic invariants $u$, $\tilde u$, or
$\varepsilon$ vanish. We will restrict attention to those discontinuities
that are picked up while we continue the kinematic invariants into
so-called Mandelstam regions, i.e. into regions in which some of the
s-like variables $s_i$ are negative, see \figref{fig:configuration}.
These  regions are reached by continuing the energies $p^0_j$ of
outgoing particles with indices $j \in I \subset \brc{4,\dots, n-1}$
to negative values. The choice of the subset $I$ labels the different
Mandelstam regions.

To each such Mandelstam region $I$ we associate an $n$-component object
$\rho^I = (\varrho^I_j)$ such that
\begin{equation}
\varrho^I_j = \left\{ \begin{array}{l}  -1 \quad \text{if} \ j \in I
\\[1mm] \  0 \quad \text{if} \ j \in \{1 \equiv n+1,2\}
\\[1mm] +1 \quad \text{otherwise}\ . \end{array} \right.
\end{equation}
In order to reach a region $\rho = (\varrho_a)$, the curve in the space
of kinematic invariants has to wind around the endpoints of some of our
branch cuts. For the points $u_{ij}=0$, the winding numbers
are~\cite{Bartels:2014ppa}
\begin{equation}
n_{ij}(\rho) = \frac{1}{4} (\varrho_{i+1}-\varrho_{i+2})
(\varrho_{j+1}-\varrho_{j}) \ .
\end{equation}
From the  formulas in~\cite{Bartels:2014ppa} one can also conclude that
the points $\varepsilon_j=0$ possess trivial winding for all Mandelstam
regions $\rho$  while for $\tilde u_j$ one finds
\begin{equation}
n_j (\rho) = \frac{1}{2}(\varrho_{j}-\varrho_{j+1})\ .
\end{equation}
With these basics set up, our next aim is to establish a number of
relations between the winding numbers for different Mandelstam
regions $\rho$. The simplest winding numbers appear for the Mandelstam
regions $\rho^{[k,l]}$ that are associated with sets $I = [k,l] = \{k,
k+1,\dots,l-1,l\}$ with $4 \leq k \leq l \leq n-1$. For these
regions, the winding numbers around the branch points in $u_{ij}$
are
\begin{equation}
n_{ij}(\rho^{[k,l]}) = \delta_{i,k-2} \delta_{j,l} \ .
\end{equation}
while for the branch points in $\tilde u_j$ one finds
\begin{equation}
n_j(\rho^{[k,l]}) = \delta_{j,k-1}-\delta_{j,l}\ .
\end{equation}
Note that any such region with $k \neq l$ is associated with a unique
cross ratio $u_{k-2,l}$ that winds during the continuation into the
Mandelstam region $\rho^{[k,l]}$. Let us point out that the above
formulas can also be applied to regions $\rho^k = \rho^{\{k\}}$
associated with a single sign flip by setting $k=l$. In this case
the winding numbers $n_{ij}$ vanish, while $n_j(\rho^k) =
\delta_{j,k-1}-\delta_{j,k}$.

The winding numbers $n_{ij}$ around the branch points $u_{ij}=0$
obey two interesting relations that will become important later
on. It is not difficult to see that
\begin{equation} \label{wone}
n_{ij} (\rho^I) = \sum_{\{k,l\} \subset I} n_{ij}(\rho^{\{k,l\}})
\end{equation}
and
\begin{equation}\label{wtwo}
n_{ij} (\rho^{\{k,l\}})  = n_{ij}(\rho^{[k,l]}) -
n_{ij}(\rho^{[k+1,l]})  -n_{ij}(\rho^{[k,l-1]}) +
n_{ij}(\rho^{[k+1,l-1]})    \ .
\end{equation}
Let us stress that these  relations are not to be read as equalities
in homology, since they do not hold for the winding numbers $n_j$.
Given the symbol for the 2-loop $n$-gluon remainder function,
\begin{equation}\label{symb}
 \smbop{R} = \sum u_{ij} \otimes S_{ij} + \sum \left(\tilde u_j \otimes S_j +
\varepsilon_j \otimes \tilde S_j\right)
\end{equation}
we can associate a symbol of length three
\begin{equation} \label{symbrho}
 \smbop{R}_\rho \equiv -2\pi i \sum n_{ij}(\rho) S_{ij} -2\pi i
\sum n_j(\rho) S_j \ .
\end{equation}
to every Mandelstam region. Here we have used that none of our curves
winds around the branch points in $\varepsilon_j =0$. Like the original
symbol, the symbols $S_{ij}$ and $S_j$ of the cut contributions are
a bit complicated. What matters to us is that they can be worked out
from the formula for $\smbop{R}$.

\section{Multi-Regge limit and relations}
\label{sec:mrlrelations}
\def\MRL{\mathrm{MRL}}

The multi-Regge limit is a scaling limit in which the pairwise
subenergies $s_{j-3} = (p_{j-1}+p_{j})^2$ for $j=4,\dots,n-1$
are sent to infinity while keeping the t-like variables in
\figref{fig:configuration} along the so-called Toller angles
finite, see~\cite{Bartels:2012gq} for details. One can show
that this limiting procedure is equivalent to sending
\begin{equation}\label{eq:eps_tozero}
\varepsilon_j \rightarrow 0
\end{equation}
while keeping $\tilde u_j$ and $(1-u_{j-2,j+1})^2\varepsilon_j^{-1}$ finite
for $j=4,\dots,n-2$. The entries $X$ of the symbol are functions
of the kinematic invariants. For each of them there exists a unique
monomial $X^\MRL$ in the variables $\varepsilon_j$ such that
\begin{equation}
\lim_{|\varepsilon_j|\rightarrow 0} X/X^{\MRL} = 1 \ .
\end{equation}
The coefficient factor in $X^{\MRL}$ still depends on $2n-10$
kinematic invariants $w_j,\bar w_j$, see below.

In order to compute the multi-Regge limit of the symbol of the
discontinuities, we parametrize the entries in terms of a natural set
of $3(n-5)$ variables similar to the ones used in the computation of
the OPE for polygon Wilson
loops~\cite{Alday:2010ku,Gaiotto:2010fk,Gaiotto:2011dt,Sever:2011pc,Basso:2013vsa},
see \appref{app:mrlparam}. For the following general discussion, it will be sufficient
to know the multi-Regge limit of the first entry. This is given
by~\cite{Bartels:2012gq}
\begin{equation}\label{eq:crossratiosMRK} u_{ij}^\MRL = 1 \quad  , \quad
\tilde u_j^\MRL  =  w_j\bar w_j\quad , \quad
\varepsilon_j^\MRL = \varepsilon_j\ .
\end{equation}
Note that we have introduced $w_a\bar w_a, a=1,\dots,n-5$ through
the multi-Regge limit of $\tilde u_j$. In order to determine the complex
variables $w_a$ and $\bar w_a$ themselves, rather than their products
only, we need to consider further combinations of kinematic invariants,
namely
\begin{equation}\label{eq:wplus1_definition}
\left[(1-u_{j-2,j+1})U_{j-1,n}^{-1}\right]^\MRL = (1+w_j)(1+\bar w_j)\ .
\end{equation}
Now that we have defined the multi-Regge limit, let us return to the
remainder function and determine the multi-Regge limit of its symbol.
Let us recall that the remainder function $R_n$ on the Euclidean sheet,
where all the $s_i > 0$, vanishes in the multi-Regge limit. This implies
that also its symbol must vanish. If we apply our construction of $S^\MRL$ to
the symbol~\eqref{symb} we find that
\begin{equation}
0 =  \smbop{R}^\MRL = \sum \left(w_j
\bar w_j \otimes S^\MRL_j + \varepsilon_j \otimes
\tilde S_j^\MRL\right) .
\end{equation}
Here we have dropped all terms containing $S_{ij}$ since $1 \otimes
S^\MRL_{ij} \equiv 0$. From the previous equation we conclude that
$S^\MRL_j = 0 = \tilde S^\MRL_j$. If we insert this into our expressions~\eqref{symbrho}
for the cut contributions we obtain\footnote{We thank James Drummond
for discussions around this point.}
\begin{equation} \label{symbrhoMRL}
 \smbop{R}^\MRL_\rho = -2\pi i \sum n_{ij}(\rho) S^\MRL_{ij}\ .
\end{equation}
Before we draw further conclusions from here, let us comment
that our result $S_j^\MRL =0$ is fully consistent with the well-known
fact that there are no non-trivial cut contributions to the continuation into
Mandelstam regions $\rho^k$ associated with a single sign
flip~\cite{Mandelstam:1963cw} (reviewed in~\cite{Bartels:2014jya}).
As we saw before, while we continue into these Mandelstam regions, only
the variables $\tilde u_j$ with $j=k,k-1$ wind around $\tilde u_j=0$. In
order for the cut contribution to cancel we must have $S_{k}^\MRL =
S_{k-1}^\MRL$, i.e.\ all these symbols must be identical, which is
consistent with the stronger statement $S_j^\MRL=0$ we derived above.

The result~\eqref{symbrhoMRL} says that the multi-Regge limit of the
symbol for any region $\rho$ only depends on the winding numbers $n_{ij}$.
Consequently, the two relations~\eqref{wone} and~\eqref{wtwo} translate
into relations for the symbol
\begin{equation}
\smbop{R_n}^{\mathrm{MRL}}_I
=
\sum_{\brc{k,l}\subset I}
\smbop{R_n}^{\mathrm{MRL}}_{\brc{k,l}}\, .
\label{eq:rel1}
\end{equation}
and
\begin{equation}
\smbop{R_n}^{\mathrm{MRL}}_{\brc{k,l}}
=
\smbop{R_n}^{\mathrm{MRL}}_{[k,l]}
-\smbop{R_n}^{\mathrm{MRL}}_{[k,l-1]}
-\smbop{R_n}^{\mathrm{MRL}}_{[k+1,l]}
+\smbop{R_n}^{\mathrm{MRL}}_{[k+1,l-1]}
\,.
\label{eq:rel2}
\end{equation}
These relations imply that we can reconstruct the symbol of all Mandelstam
regions from the symbol of those $(n-4)(n-5)/2$ regions that are associated
with (any number of) adjacent flips. Let us note that the relations
in this and the preceding section are independent of the loop order, as long
as we restrict to the contributions of maximal functional transcendentality.
In the case of $n=7$ external gluons, such relations between different
multi-Regge regions were first explored in~\cite{Bartels:2014jya}, see
e.g.\ formulas (6.1)-(6.8) in the concluding section of that paper. The
last formula of that list, for instance, correspondents to our equation~\eqref{eq:rel2}
with $n=7, k=4$ and $l=6$, keeping in mind that
$\smbop{R_n}^{\mathrm{MRL}}_{[5,5]}$ vanishes. Our results extend such
relations to arbitary numbers of gluons, at least for the terms of
maximal functional transcendentality which are captured by the symbol.

\section{Building blocks of the symbol}
\label{sec:blocks}

Having found all these relations between the Mandelstam regions
we want to finally describe the multi-Regge limit of the symbol
for the regions $\rho^{[k,l]}$, in which the adjacent particles
$k,k+1,\ldots,l$ have their energy signs flipped. From these
symbols, we will be able to construct the symbols in
all other regions $\rho = \rho^I$ as linear combinations,
following the relations~\eqref{eq:rel1} and~\eqref{eq:rel2}. For
the regions $\rho^{[k,l]}$, we find
\begin{equation} \label{eq:Rnkl}
\frac{\smbop{R_n}^{\mathrm{MRL}}_{[k,l]}}{2\pi i}
=
\sum_{i=k}^{l-1}\left(f(v_{i})\log\varepsilon_{i} +
\tilde f(v_{i})\right) + \sum_{i=k}^{l-2}g(v_{i},v_{i+1})\,,
\end{equation}
where the terms in the first sum are obtained from the two-loop remainder function for $n=6$
external gluons as
\begin{equation}\label{fsymboldef}
 f(v_4)\log\varepsilon_4+\tilde f(v_4)=
\frac{\smbop{R_6}^{\mathrm{MRL}}_{[4,5]}}{2\pi i}\,,
\end{equation}
and the symbol $g$ that appears in the second sum is related
to the two-loop remainder function for $n=7$ external gluons
through
\begin{equation}
g(v_4,v_5)=
\frac{\smbop{R_7}^{\mathrm{MRL}}_{[4,6]}}{2\pi i}-\sum_{i=4,5} \left(f(v_{i})\log\varepsilon_i  +
\tilde f(v_i)\right) \,.\label{gsymboldef}
\end{equation}
In order to recycle this data from $n=6$ and $n=7$ for the symbol of the
two-loop remainder function for any number $n$ of external gluons, as
described in equation~\eqref{eq:Rnkl}, we introduced a new set of variables
$v_i, i=k,\dots,l-1$ that are  related to the kinematic variables
$w_i$ by
\begin{equation}\label{wtov}
w_j=\frac{(v_j-v_{j-1})(1+v_{j+1})}{(v_{j+1}-v_j)(1+v_{j-1})}\,,
\quad
j\in\brc{k,\dots,l-1}\,,
\end{equation}
with the boundary conditions $v_{k-1}=0$, $v_{l}=\infty$. Let us stress
that this map between the $v_i$ and $w_j$ depends on the Mandelstam region
$\rho^{[k,l]}$ we consider, and let us mention a few examples for concreteness: In the $\rho^{[4,5]}$ region relevant for $R_6$ in (\ref{fsymboldef}) we simply have $w_4\equiv w=v_4$, whereas in the $\rho^{[4,6]}$ region relevant for $R_7$ in (\ref{gsymboldef}) we have
\begin{equation}
w_4= \frac{v_4 \left(1+v_5\right)}{v_5-v_4}\,,\quad\qquad w_5= \frac{v_5-v_4}{1+v_4}\,.
\end{equation}
For completeness, the inversion of (\ref{wtov}) reads
\begin{equation}
v_j=\frac{%
(1+(1+(\dots(1+w_k)w_{k+1})\dots)w_{j-1})w_jw_{j+1}\dots w_{l-1}%
}{%
1+(1+(\dots(1+w_{j+1})w_{j+2})\dots)w_{l-1}%
}\,,
\quad
j\in\brc{k,\dots,l-1}\,.
\end{equation}

In order to fully describe the symbol for all Mandelstam regions, it remains
to spell out formulas for the symbols $f$, $\tilde f$ and $g$. The
expression for $f$ is known~\cite{Prygarin:2011gd}, it reads
\begin{equation}\label{fsymbol}
f(w)=
\frac{1}{2}\biggsmb{(1+w)(1+\bar w)\otimes \frac{(1+w)(1+\bar w)}{w\bar w}}+
\frac{1}{2}\biggsmb{\frac{(1+w)(1+\bar w)}{w\bar w} \otimes (1+w)(1+\bar w)}\ .
\end{equation}
Similar expressions for $\tilde f$  and $g$ can be found in the \mathematica\
file \texttt{SR2MRL.m} accompanying this publication. Let us only mention that the letters of the
symbol $g(-1/y,-x)$
are $x$, $y$, $(1-x)$, $(1-y)$, $(1-x y)$, and their complex conjugates. Leaving
the complex conjugate letters aside for a moment, functions with this five-letter
symbol alphabet belong to the class of 2-dimensional harmonic polylogarithms
(2dHPLs)~\cite{Gehrmann:2000zt}, and in particular to the same subset
which was found to describe the contribution of all single-particle
gluonic bound states in the OPE expansion of the six-point remainder
function~\cite{Drummond:2015jea}.

Before concluding this section, let us summarize how the
result~\eqref{eq:Rnkl}-\eqref{wtov} was obtained. Our starting point,
the $n$-point two-loop MHV symbol in general kinematics has been
derived in~\cite{CaronHuot:2011ky}, by extending the duality between
MHV amplitudes and bosonic Wilson
loops~\cite{Alday:2007hr,Drummond:2007aua,Brandhuber:2007yx} to the
supersymmetric case. The result, which is also contained in a
\mathematica\ file accompanying the \texttt{arXiv} submission of the
aforementioned article, encodes the kinematical dependence through
momentum twistors $Z_i$, and in particular their scalar products,
known as four-brackets,
\begin{equation}
\br{Z_iZ_jZ_kZ_\ell}\equiv \epsilon_{abcd} Z_i^aZ_j^bZ_k^cZ_{\ell}^d\equiv\br{ijkl}\,,
\end{equation}
and their bilinears
\begin{align}
\br{ij(abc)\cap(def)}&\equiv\br{iabc}\br{jdef}-\br{jabc}\br{idef}\,,\\
\br{i(ab)(cd)(ef)}&\equiv\br{aicd} \br{bief}-\br{aief} \br{bicd}\,.
\end{align}
We evaluate these expressions in any convenient set of variables, for
example the one described in \appref{app:mrlparam}, and Taylor expand
the symbol entries around the multi-Regge limit (in this case
$T_i\to0$, with $S_i/T_i$ fixed, for $i=1,\ldots, n-5$), keeping only
the first term in the expansion of each entry. We may then trade the
expansion parameters for the kinematic parameters $\varepsilon_i$ we
defined in eq.~\eqref{eq:ut_eps_crossratios} that also become small
in multi-Regge kinematics, see eq.~\eqref{eq:eps_tozero}, and the surviving
$2(n-5)$ kinematical parameters for the $w_i, \bar w_i$ variables
defined in eqs.~\eqref{eq:crossratiosMRK,eq:wplus1_definition}.
After that, we factor the symbol entries and then expand the factors
according to the symbol property,
\begin{equation}\label{eq:expand_symbol}
A\otimes (X_a X_b)\otimes B=A\otimes X_a \otimes B+A\otimes X_b \otimes B\,,
\end{equation}
which in particular also implies
\begin{equation}
A\otimes X^m\otimes B=m\,(A\otimes X \otimes B)\,.
\end{equation}
These stem from the definition~\eqref{eq:iteratedint,eq:symbol}, where
it is evident that the symbol behaves as a tensor product of
differentials of logarithms. Thus it also obeys the property
\begin{equation}
A\otimes c \otimes B=0
\end{equation}
for any nonzero constant $c$, allowing us to discard terms of this form.

Furthermore, we need to extract the divergent logarithms in the variables
$\varepsilon_i$. In general, the structure of the limit is such that,
at loop order $L$, divergences of degree $\log^{L-1}$ appear. At two loops, we
will thus have at most single logarithms, which come from symbols with
one of the $\varepsilon_i$ in one entry. We may extract the divergent
logarithms by virtue of the shuffle identity
\begin{equation}
\varepsilon_i\otimes X\otimes Y
=
\log\varepsilon_i\,\smb{X\otimes Y}-\smb{X\otimes \varepsilon_i\otimes Y}-\smb{X\otimes Y\otimes \varepsilon_i}\,,
\end{equation}
which follows from writing $\log\varepsilon_i$ as an iterated integral
and nesting its integration range with the $(X\otimes Y)$ integral it
multiplies.

Finally, we have worked out the transformation~\eqref{wtov}, relating
the multi-Regge variables $w_i$, $\bar w_i$ most commonly defined in
terms of the cross ratios~\eqref{eq:crossratiosMRK,eq:wplus1_definition},
to a generalization of the variables used previously~\cite{Prygarin:2011gd,Bartels:2011ge}.
Expanding once more the different factors in the symbol entries as in eq.~\eqref{eq:expand_symbol},
we observe the structure~\eqref{eq:Rnkl} up to $n=10$ points, and
conjecture it to hold for all multiplicities.

\section{From symbols to functions}
\label{sec:function}

Of course it is of interest to lift the symbols $f,\tilde f$  and $g$
to functions. For $f$ and $\tilde f$ the answer is well-known even beyond
the NLLA since the six gluon remainder function is known explicitly to
very high loop order in multi-Regge
kinematics~\cite{Dixon:2012yy,Drummond:2015jea,Broedel:2015nfp}, and
implicitly also to all orders~\cite{Basso:2014pla,Basso:2015uxa}. For
completeness, let us quote here the relevant two-loop result
\cite{Lipatov:2010ad} in our conventions (\ref{fsymboldef}), where by
slight abuse of notation $f$, $\tilde f$ now denote functions rather than
symbols,%
\footnote{Our expressions have an extra factor of 4 compared to
\cite{Lipatov:2010ad} due to the use of $\lambda/(4\pi)^2$ as
expansion parameter, as in \cite{CaronHuot:2011ky}. In addition, here
we have replaced the large logarithm by $\log (1-u_1)\to
\frac{1}{2}\log \varepsilon_4-\frac{1}{2}\log \frac{|w|^2}{|1+w|^4}$.}
\begin{align}
f(w)&=\frac{1}{2} \log |1+w|^2 \log \frac{|1+w|^2}{|w|^2}\,,\nn\\
\tilde f(w)&=-4 \text{Li}_3(-w)-4 \text{Li}_3(-\bar w)+2 \log |w|^2 (\text{Li}_2(-w)+\text{Li}_2(-\bar w))\\
&\mspace{21mu}+\frac{1}{3} \log ^2|1+w|^2 \log \frac{|w|^6}{|1+w|^4}-\frac{1}{2} \log |1+w|^2 \log \frac{|1+w|^2}{|w|^2} \log \frac{|w|^2}{|1+w|^4}\,,\nn
\end{align}
with $|w|^2=w\bar w$ and $|1+w|^2=(1+w)(1+\bar w)$.

For the function associated to $g$, while it could in principle be
obtained from the formula for the two-loop heptagon remainder function
presented in~\cite{Golden:2014xqf}, the latter is only valid in a
subspace of the Euclidean region known as the positive region, thus
rendering the analytic continuation relevant for the multi-Regge limit
quite intricate.

Instead, in this section we will construct a prototype function whose
symbol equals $g$ by directly comparing it against a basis
of functions having the same alphabet. That is, we first construct an
ansatz for the function, which is a linear combination with
undetermined coefficients, of all independent functions of
transcendentality $m=0,1,2,3$, multiplied by transcendental constants
such as $(i\pi)^k$ or (multiple) zeta values $\zeta_k$, so that each
term has uniform total transcendentality $m+k=3$. Equating the symbol
of the ansatz with $g$ then fixes the coefficients of all terms with
$m=3$. We further reduce the ambiguity of the remaining terms with
lower functional transcendentality by imposing simple constraints from
symmetry, ending up with a prototype function with 25 undetermined
parameters.

We start by noting that a basis spanning the subset of 2dHPLs with the
five-letter (unbarred) alphabet mentioned below~\eqref{fsymbol} can be
generated from
\begin{equation}\label{2dHPLbasis}
\mathcal{G}=\Big\{G(\vec a;y)|a_i\in \{0,1\}\Big\} \cup
\Big\{G(\vec a;x)|a_i\in \{0,1,1/y\}\Big\}\,,
\end{equation}
where
\begin{equation}
\label{G_def}
G(a_1,\ldots,a_n;z)
\equiv
\begin{cases}
\frac{1}{n!}\log^n z & \text{if }a_1=\ldots=a_n=0\,\\[1ex]
\int_0^z \frac{dt_1}{t_1-a_1}G(a_2,\ldots,a_n;t_1) & \text{otherwise,}
\end{cases}
\end{equation}
with $G(;z)=1$, are iterated integrals over a particular curve, known
as Goncharov or multiple polylogarithms (MPLs). The basis~\eqref{2dHPLbasis}
is in turn the part of the hexagon function basis
considered in~\cite{Dixon:2013eka} (before imposing branch cut
conditions) that is independent of one of the
three so-called $y$-variables. From the recursive definition of the
symbol of MPLs,
\begin{multline}\label{Gsymbol}
\smbop{G(a_{n-1},\ldots,a_1;a_n)}=
\sum_{i=1}^{n-1} \Big[
 \smbop{G(a_{n-1},\ldots,\hat a_i,\ldots, a_1;a_n)}\otimes (a_i-a_{i+1})\\
-\smbop{G(a_{n-1},\ldots,\hat a_i,\ldots,a_1;a_n)}\otimes (a_i-a_{i-1})\Big]\,,
\end{multline}
where $a_0=0$ and hatted indices are omitted, it is straightforward to
see that the 2dHPLs in eq.~\eqref{2dHPLbasis} indeed yield the five-letter
alphabet mentioned below eq.~\eqref{fsymbol}.

In fact, allowing the entries of the singularity vector $\vec a\equiv
(a_1,\ldots,a_n)$ to take any value within the prescribed set in
the basis~\eqref{2dHPLbasis} yields an overcomplete system, because
of shuffle identities such as
\begin{equation}
G(a;z)\,G(b;z)=G(a,b;z)+G(b,a;z),
\end{equation}
which follow from the definition~\eqref{G_def} by nesting the
integration range of the integrals on the left-hand side. According to
Radford's theorem~\cite{RADFORD1979432}, we may solve these identities
and obtain a linearly independent set of functions by only keeping the
singularity vectors that form Lyndon words. That is, if we consider
all words made of letters of a given alphabet, the latter also
defining a particular ordering between the letters, then Lyndon words
are those words that no matter how we split them into two substrings,
the left substring is always lexicographically smaller than the right
substring.

For example, all Lyndon words up to length three of the alphabet $0<1<2$ are
\begin{equation}
0,1,2,01,02,12,001,002,011,012,021,022,112,122,
\end{equation}
and from this example we may obtain all irreducible 2dHPLs
of eq.~\eqref{2dHPLbasis} as follows:%
\footnote{Namely, those 2dHPLs which cannot be
written as a product of lower-weight 2dHPLs.}
Replacing
$2\rightarrow1/y$ yields the singularity vectors $\vec a$ of all
irreducible 2dHPLs on the right hand side of eq.~\eqref{2dHPLbasis}, and
discarding all words with the letter $2$ yields the respective ones on
the left-hand side of the latter formula. Let us call the Lyndon
basis of $G(\vec a;y), G(\vec a;x)$, with singularity vectors as
obtained by the aforementioned two operations, as
$\mathcal{G}^L\subset \mathcal{G}$.

So far we have constructed irreducible functions with only half of the
ten-letter alphabet appearing in the symbol $g$. Clearly, functions
for the other half of the alphabet may be obtained by the complex
conjugate of $\mathcal{G}^L$, $\overline{\mathcal{G}}^L$. Now it turns
out that as a consequence of the local path independence of iterated
integrals, also known as the integrability condition, and the fact
that no letter mixes barred and unbarred variables (for example, we
don't encounter letters of the form $1+w \bar w$), there exist no
other irreducible functions whose symbol entries span the entire
ten-letter alphabet in question.

In summary, all irreducible functions with the same alphabet as $g$
are given by $\mathcal{G}^L\cup\overline{\mathcal{G}}^L$, and to
obtain a complete basis at a given weight, one needs to add to the
latter all distinct products of lower-weight functions from the same
set, and with the same total weight. In this manner, we obtain a basis
of dimension 1, 10, 63, and 320 at weights 0 (i.e.\ $G[;z]=1$), 1, 2, and 3, respectively.
Forming a linear combination of functions at weight 3 with arbitrary
coefficients, taking its symbol with the help of eq.~\eqref{Gsymbol}
and equating the result with the symbol $g$, we uniquely determine
the 320 coefficients. In particular, we find that (exceptionally, $g$ here denotes the function rather than the symbol)
\begin{equation}
\begin{aligned}
2g(-1/y,-x)=&G(1,x) G(1,y) G\left(0,\bar{x}\right)+G(1,x) G(1,y) G\left(0,\bar{y}\right)+G(0,x) G(1,y) G\left(1,\bar{x}\right)\\
&+G(0,y) G(1,y) G\left(1,\bar{x}\right)-2 G(1,x) G(1,y) G\left(1,\bar{x}\right)\\
&-G(1,y) G\left(0,\bar{y}\right) G\left(1/y,x\right)+G(0,y) G(1,y) G\left(1/\bar{y},\bar{x}\right)\\
&+G(0,x) G(1,x) G\left(1,\bar{y}\right)+G(1,x) G(0,y) G\left(1,\bar{y}\right)\\
&-G(1,x) G\left(0,\bar{x}\right) G\left(1/y,x\right)+G(0,x) G\left(1,\bar{x}\right) G\left(1/y,x\right)\\
&+G(0,y) G\left(1,\bar{y}\right) G\left(1/y,x\right)+G(0,x) G(1,x) G\left(1/\bar{y},\bar{x}\right)\\
&-2 G(0,1,x) G\left(1/\bar{y},\bar{x}\right)-2 G(0,1,y) G\left(1/\bar{y},\bar{x}\right)-2 G\left(1,\bar{x}\right) G\left(0,1/y,x\right)\\
&+2 G\left(1,\bar{x}\right) G\left(1,1/y,x\right)-2 G\left(1,\bar{y}\right) G\left(1,1/y,x\right)-2 G(0,x) G(1,x) G\left(1,\bar{x}\right)\\
&+2 G(0,1,x) G\left(0,\bar{x}\right)+2 G(0,1,x) G\left(1,\bar{x}\right)-2 G(0,y) G(1,y) G\left(1,\bar{y}\right)\\
&+2 G(0,1,y) G\left(0,\bar{y}\right)+2 G(0,1,y) G\left(1,\bar{y}\right)+G(0,x) G(1,x) G(1,y)\\
&+G(1,x) G(0,y) G(1,y)-G(0,y) G(1,y) G\left(1/y,x\right)-2 G(1,y) G\left(1,1/y,x\right)\\
&-G(0,x) G(1,x) G\left(1/y,x\right)+2 G(0,1,x) G\left(1/y,x\right)+2 G(0,1,y) G\left(1/y,x\right)\\
&+2 G(1,x) G\left(0,1/y,x\right)+2 G(1,x) G\left(1,1/y,x\right)-4 G\left(0,1,1/y,x\right)\\
&-4 G\left(0,1/y,1,x\right)-4 G\left(1,1,1/y,x\right)\\
&+2 G(0,x) G(0,1,x)-2 G(1,x) G(0,1,x)-4 G(0,0,1,x)+4 G(0,1,1,x)\\
&-2 G(0,1,y) G(1,y)+2 G(0,y) G(0,1,y)-4 G(0,0,1,y)+4 G(0,1,1,y)\\
&+ (x\leftrightarrow\bar{x}, \,y\leftrightarrow\bar{y}) + \text{lower-weight functions}\,.
\end{aligned}
\end{equation}
In order to fix the function completely, we
need to address the coefficients of the remaining lower-weight
functions multiplying the independent transcendental constants
$(i\pi), \zeta_2, (i\pi)^3$ and $\zeta_3$, so that the total weight is
3, which thus sum up to 75.

We have explained before that this information on contributions of
lower weight is invisible to the symbol. However we may further
constrain these terms by examining the symmetries of the problem. MHV
amplitudes are invariant under parity (spatial reflection), which in
the multi-Regge limit amounts to the transformation
$w_i\leftrightarrow \bar w_i$~\cite{Prygarin:2011gd}. Imposing this
condition on our ansatz leaves 2, 5, and 34 undetermined parameters
multiplying functions of weight 0, 1, and 2, respectively.

Furthermore, we have target-projectile symmetry, corresponding to an
invariance under exchange of the incoming momenta, or
equivalently under $w_k\to 1/w_{n-4-k}$, see e.g.~\cite{Bartels:2014ppa}.
For $g(v_1,v_2)$ as defined in eqs.~\eqref{gsymboldef}-\eqref{wtov}, and given
that $f(v)=f(1/v)$~\cite{Prygarin:2011gd} and similarly for $\tilde f$, this amounts to symmetry
under $v_1\leftrightarrow 1/v_2$ (or $x\leftrightarrow y$), which further reduces the number of
unknowns to 2, 3, and 20, by order of increasing functional transcendentality.
The only additional subtlety when imposing this symmetry comes from mapping
the transformed functions back into the basis. For all but one, this can be
done immediately due to the following property of MPLs,
\begin{equation}\label{Grescale}
G(a_1,\dots,a_k;z)
=
G(x a_1,\dots,x a_k; x z)\,.
\end{equation}
for $a_k\ne0$ and $x\in\mathbb{C}^*$. And for the one left, we may use the
identity
\begin{equation}
G(1,1/x;y)=G(1;x) G(1,y)+G(0,1/y;x)-G(1,1/y;x)\,,
\end{equation}
which follows from the quasi-shuffle algebra of MPLs, see for
example~\cite{Duhr:2014woa}.

This concludes the discussion on the use of symmetry in constraining the terms of
lower functional transcendentality. The final result for the
25-parameter functional representative for $g$ (also including the functions for $f,\tilde f$), or equivalently the
two-loop seven-point remainder function in multi-Regge kinematics,
is included in the \mathematica\ file \texttt{gfunction.m} attached to this publication.
It would be interesting to fix the function completely, by further
exploiting its expected analytic properties~\cite{Dixon:2012yy},
overlap with the collinear
limit~\cite{Basso:2014pla,Drummond:2015jea}, or better yet by
constructing the seven-point remainder function with proper branch
cuts in general kinematics, and taking its limit. We leave these
exciting questions for future work.

\section{Conclusions}
\label{sec:conclusions}

Let us comment a bit more on the three main results of this
paper. At the end of \secref{sec:mrlrelations} we found a set of relations that
determine the multi-Regge limit of the symbol of all Mandelstam
regions from the regions $I=[k,l]$ with adjacent flips. The
argument we presented is actually not restricted to the two-loop
remainder function; in fact, it can easily be seen that it extends
to all loops. This does not imply, however, that the multi-Regge
limit of the remainder function itself satisfies similar relations.
While the relations hold for the terms of maximal functional
transcendentality, they are well known to receive additional
contributions from lower transcendentality, such as double cut
contributions. It would be interesting to study these modifications
in more detail.

Thereby, one should also be able to resolve the observed discrepancy
between the weak and strong coupling results for $n=7$ external gluons in
the Mandelstam region ${4,6}$ in which the signs are flipped for the outgoing
particles in position $4$ and $6$. In this case, the multi-Regge limit of the
remainder function was shown to be non-trivial at  weak coupling, in full agreement
with our analysis of the two loop symbol, while the continuation at strong
coupling produced a vanishing result~\cite{Bartels:2014ppa}. Our
investigation of the symbol suggests that the issue is related to the
choice of the curve along which the kinematic invariants are continued
in the strongly coupled theory. The curve selected in~\cite{Bartels:2014ppa}
possesses the desired winding numbers around $u_{ij}=0$ but does not seem
to belong to the right homotopy class. This issue certainly deserves
further attention.

The second outcome of our analysis concerns the building blocks
$f$ and $g$ of the multi-Regge limit. We found that, in multi-Regge
kinematics, the symbol of the two loop remainder function can be
built from terms that are entirely determined from the expressions
with $n=6$ and $n=7$ external gluons. This result is a consequence
of the low loop order. Generalizing arguments that were presented
in~\cite{Bartels:2011ge}, one can show that the multi-Regge limit
of the $L$-loop remainder function is determined by $L$ different
building blocks $g^{(\nu)},\nu=1,\dots,L$. The first two of these
are $g^{(1)}\equiv f$ and $g^{(2)}=g$. The remaining ones may be
reconstructed from processes involving up to $n=L+5$ external
gluons. They receive their leading contribution at N$^{\nu-1}$LLA.
In going to higher loop orders, the building blocks $g^{(\nu)}$
themselves pick up higher order terms from the expansion in
large logarithms. For $g^{(1)}=f$, for example, our two loop
analysis only allowed to determine LLA and NLLA terms. In order
to find $g^{(1)}$ to N$^2$LLA accuracy, we need to analyze the
known three loop symbol for $n=6$ external gluons.

Let us finally mention that our results also impose strong
constraints on the production vertex that appears in the multi-Regge
limit for $n=7$ external gluons. In particular, by transforming the
prototype function for $g$, we could fix this vertex in NLLA, up to 25
parameters. In LLA, the relevant production vertex was actually computed
by Bartels et al.~\cite{Bartels:2014jya}. It would be interesting to reproduce
their result from our expressions, and to extract its
NLLA corrections. This could in turn be used as a seed
in order to compute the $n=7$ remainder function to NLLA in principle
at any loop order from the BFKL formula of~\cite{Bartels:2014jya},
thus providing potentially useful boundary data for the amplitude
bootstrap
program~\cite{Dixon:2011pw,Dixon:2013eka,Dixon:2014xca,Drummond:2014ffa,Dixon:2014iba,Dixon:2015iva}.
More generally, the
integrability of $\superN=4$ SYM theory raises the hope that all
power-suppressed terms can be obtained to all loops also for any
 of points, similarly to the $n=6$ case~\cite{Basso:2014pla}.

\subsection*{Acknowledgments}

We wish to thank Jochen Bartels, Lance Dixon, James Drummond, Mark
Spradlin, and Martin Sprenger for
useful discussions and comments.
The work of T.\,B.~is supported by a Marie Curie International Outgoing
Fellowship within the 7$^{\mathrm{th}}$ European Community Framework
Programme under Grant No.~PIOF-GA-2011-299865. The research leading to
these results has received funding from the US Department of
Energy under contract DE-AC02-76SF00515, the People Programme (Marie
Curie Actions) of the European Union's Seventh Framework Programme
FP7/2007-2013/ under REA Grant Agreement No 317089 (GATIS), and from
the SFB 676 ``Particles, Strings and the Early Universe''.

\appendix

\section{Parametrization of the Multi-Regge Limit}
\label{app:mrlparam}

We seek a good parametrization of the multi-Regge limit of (dual)
conformally inequivalent null polygons. Natural variables for
such polygons arose in the construction of the OPE for null polygon
Wilson loops~\cite{Alday:2010ku,Gaiotto:2010fk,Gaiotto:2011dt,Sever:2011pc,Basso:2013vsa}.
We will describe a slightly modified
parametrization that features a canonical multi-Regge limit for any
number of edges.

Just as for the Wilson loop OPE, we will start
an arbitrary fixed reference null $n$-gon that we tessellate into a
sequence of $(n-5)$ internal null
tetragons and two boundary tetragons. Each null tetragon is stabilized
by three conformal transformations. For each internal tetragon, we act
with its stabilizing conformal transformations on all cusps of the
polygon that lie ``below'' the internal tetragon. In this way, we
generate a $3(n-5)$-dimensional family of conformally inequivalent
null polygons. Our parametrization differs from the parametrization
for the Wilson loop
OPE~\cite{Alday:2010ku,Gaiotto:2010fk,Gaiotto:2011dt,Sever:2011pc,Basso:2013vsa} by
the specific choice of tessellation, which in our case is tailored to the
multi-Regge limit.

\paragraph{Momentum Twistors.}

The two-loop symbol for MHV amplitudes~\cite{CaronHuot:2011ky} is
expressed in terms of conformally invariant combinations of momentum
twistors~\cite{Hodges:2009hk}.
We therefore need a good parametrization of the $n$ momentum twistors
that parametrize the null $n$-gon (or $n$-point amplitude). Let us
note some key properties of momentum twistors that will be relevant in
the following.
Points $x$ in dual spacetime $\Reals^{1,3}$ are in one-to-one
correspondence with null rays $X\in\Reals^{2,4}$, $X^2=0$, $tX\cong X$~\cite{Dirac:1963ta}.
When written as a bispinor, a null vector $X$ decomposes into a pair of
spinors, $X^{ab}=Z^{[a}\tilde Z^{b]}$.
This is the usual map between spacetime points $x$ and lines
$(Z,\tilde Z)$ in twistor space.
Points $x_i$ that are null separated translate to lines in twistor
space that intersect and
thus share a common spinor (twistor). A null polygon with $n$ cusps
$x_i$
is hence parametrized by $n$ momentum
twistors $Z_i$, with $x_i\simeq(Z_i,Z_{i+1})$.
Points that lie on a common null line in spacetime map to lines in
twistor space that lie in a common plane and intersect in a common
point.
By definition,
momentum twistors are $\grp{SO}(2,4)$ spinors, hence conformal
transformations act on them via (right) multiplication by $\grp{SL}(4)$ matrices.

\paragraph{The Hexagon.}

\begin{figure}\centering
\autoparbox{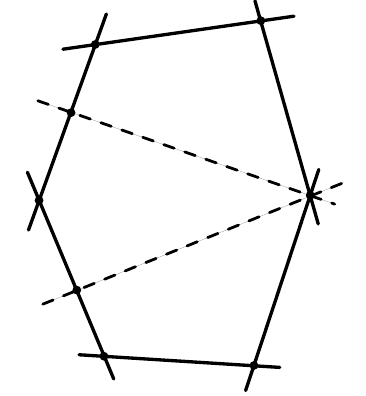}
\;\;$\longleftrightarrow$\quad
\autoparbox{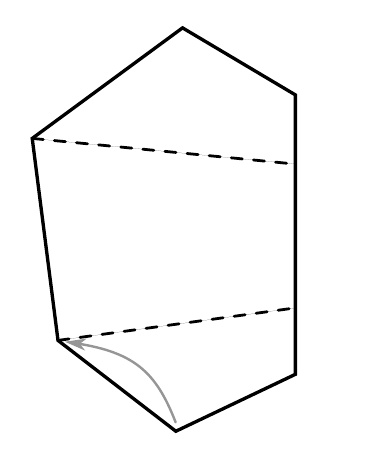}
\;$\xrightarrow[S/T\,\text{fixed}]{T\to0}$\;
\autoparbox{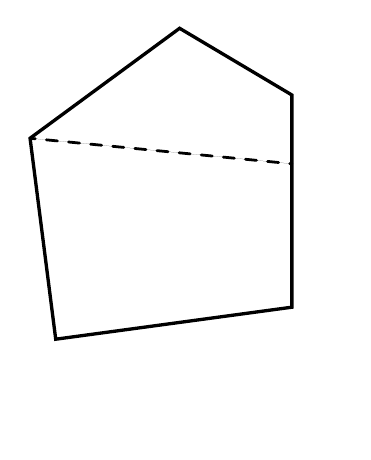}
\caption{Illustration of the hexagon tessellation (center), its
momentum-twistor picture (left), and the $2\to4$ Regge limit (center
to right).}
\label{fig:hexagon}
\end{figure}

In a conformal theory, the simplest non-trivial null polygon is
the hexagon. All null tetragons are conformally equivalent;
the same is true for null pentagons. To parametrize an arbitrary fixed
reference hexagon, pick six numerical momentum twistors
$Z_{1,\dots,6}$. Each line $(Z_i,Z_{i+1})$ defines a cusp $x_i$ of the
hexagon. Now tessellate the hexagon by
drawing the two (unique) null lines that connect the cusps
$x_4$ and $x_5$ with the line $x_{12}$. These lines are characterized
by points $x\supup{int}_4$, $x\supup{int}_5$ on the line $x_{12}$, or
equivalently by the momentum twistors $Z\supup{int}_{4}$,
$Z\supup{int}_5$ that mark the intersections of the lines
$(Z_4,Z_5)$ and $(Z_5,Z_6)$ with the plane $(Z_1,Z_2,Z_3)$, see
\figref{fig:hexagon}.
The cusps $(x_4,x_5,x\supup{int}_5,x\supup{int}_4)$, or equivalently
the momentum twistors $(Z_5,Z\supup{int}_5,Z_2,Z\supup{int}_4)$ define
the internal null tetragon. It is preserved by three conformal
transformations~\cite{Alday:2010ku}: One rotation in the plane orthogonal to the tetragon,
parametrized by $\phi$, and two non-compact conformal transformations
parametrized by $\sigma$ and $\tau$ that move points along the
horizontal and vertical directions of the tetragon.
In momentum-twistor space, these transformations are represented by an $\grp{SL}(4)$ matrix
$M\supup{int}(F,S,T)$ that depends on the three parameters
$F=e^{i\phi}$, $S=e^{\sigma}$, $T=e^{-\tau}$.
It must preserve the four momentum twistors
of the internal tetragon, and hence is uniquely defined by its
eigenvalues:
\begin{equation}
\bigbrk{Z_5,Z_2,Z\supup{int}_5,Z\supup{int}_4}.M\supup{int}(F,S,T)
=
\sqrt{F}\diag\bigbrk{1/FS,S/F,T,1/T}.\bigbrk{Z_5,Z_2,Z\supup{int}_5,Z\supup{int}_4}
\,.
\end{equation}
A family of conformally inequivalent null hexagons
is now obtained by acting with the stabilizing matrix
$M\supup{int}(F,S,T)$
of the internal tetragon on the momentum twistors $Z_6$, $Z_1$ that
parametrize points ``below'' the internal tetragon.

The $2\to4$ multi-Regge limit of the hexagon is characterized by the following
behavior of the three independent cross ratios:
\begin{equation}
U_{36}\to0\,,
\qquad
U_{41}/U_{36}\quad\text{and}\quad
\brk{1-U_{25}}/U_{36}\quad\text{finite}\,.
\end{equation}
In our parametrization, this
limit is attained for $T\to0$ with $S/T$ fixed,%
\footnote{This is essentially the same limit as for the Wilson loop
OPE parametrization~\cite{Dixon:2013eka,Hatsuda:2014oza}.}
which can be understood as follows: The limit $T\to0$ ``flattens'' the bottom
of the hexagon, while the simultaneous limit $S\to0$ moves the bottom
cusp $x_6$ towards $x_5$, see \figref{fig:hexagon}. It is then clear
that $U_{25}$ approaches $1$, while $x_{46}$ and $x_{51}$ become
lightlike, and therefore $U_{36}$ and $U_{41}$ go to zero (all at the
same rate). In the
limit, we find the following relations between the tessellation parameters $F$,
$S/T$ and the kinematic parameters $w_4$, $\bar w_4$ introduced in
\secref{sec:mrlrelations}:
\begin{equation}
F^2=b\frac{\bar w_4}{w_4}\,,
\qquad
\frac{S^2}{T^2}=\frac{c}{w_4\bar w_4}\,.
\end{equation}
Up to $\order{T^4}$, the three independent cross ratios then expand to
\begin{equation}
U_{25}=1-\frac{(1+w_4)(1+\bar w_4)}{w_4\bar w_4}a\,T^2\,,
\qquad
U_{36}=\frac{a\,T^2}{w_4\bar w_4}\,,
\qquad
U_{41}=a\,T^2\,.
\end{equation}
The coefficients $a$, $b$, $c$ are numerical constants whose values
depend on the choice of reference hexagon.
Below, we will give an explicit example for which $a=1/4$, $b=1$, $c=-1$.

\paragraph{General Polygons.}

\begin{figure}\centering
\autoparbox{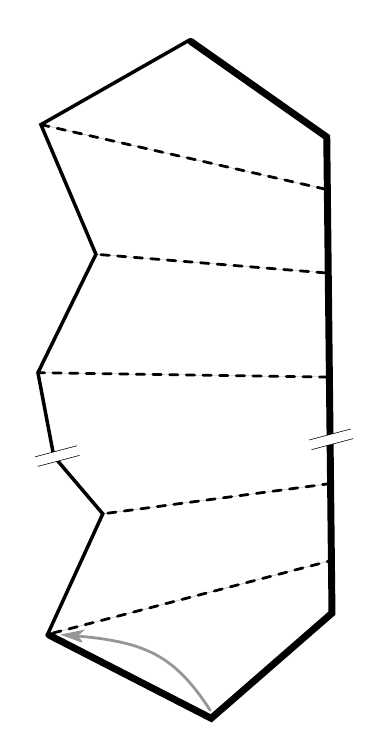}
$\xrightarrow[\frac{S_{n-5}}{T_{n-5}}\text{ fixed}]{T_{n-5}\to\,0}$
\autoparbox{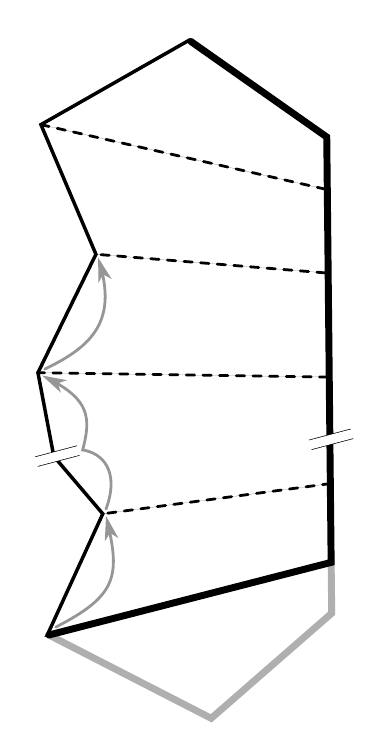}
$\xrightarrow[S_i/T_i\text{ fixed}]{T_i\to\,0\;\forall i}$
\autoparbox{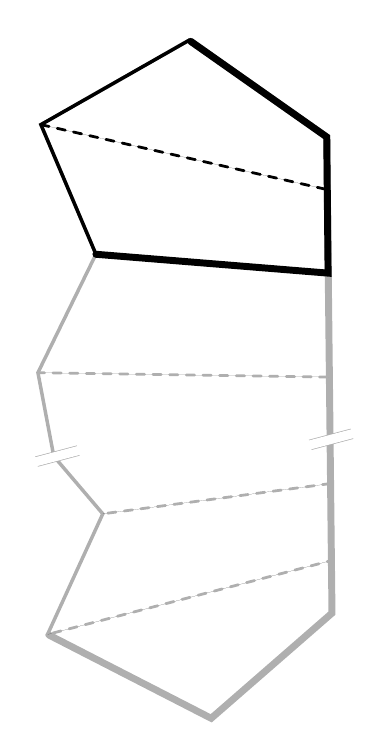}
\caption{Tessellation of the null $n$-gon that is tailored to the
$2\to(n-2)$ multi-Regge limit (left), and systematics of the
multi-Regge limit in the tessellation parameters (center and right).
When $T_{n-5}\to0$, $S_{n-5}/T_{n-5}$ fixed (center), one can see that
$U_{j,n-1}\to1$ for $j=2,\dots,n-4$, and $U_{n-3,n},U_{n-2,1}\to0$.
Similarly, $U_{i+2,n},U_{i+3,1}\to0$,
$U_{j,i+4}\to1$ for $j=2,\dots,i+1$ when $T_i\to0$, $S_i/T_i$ fixed.
The right figure shows the full multi-Regge limit.}
\label{fig:tessellation}
\end{figure}

The parametrization of the hexagon straightforwardly generalizes to
null polygons with any number of edges. Let $Z_{1,\dots,n}$ be
numerical momentum twistors that parametrize an arbitrary reference
null $n$-gon. Tessellate the $n$-gon into $(n-5)$ internal tetragons
and two boundary tetragons by drawing the (unique) null lines from
cusps $x_4,\dots,x_{n-1}$ to line $x_{12}$, see
\figref{fig:tessellation}.
These internal null lines
are characterized by intersection points $x\supup{int}_{4,\dots,n-1}$ on line
$x_{12}$, or equivalently by the momentum twistors
$Z\supup{int}_{4,\dots,n-1}$ that mark the
intersections of the twistor lines $(Z_4,Z_5),\dots,(Z_{n-1},Z_n)$ with
the twistor plane $(Z_1,Z_2,Z_3)$. Each internal tetragon is again
stabilized by three conformal transformations parametrized by
$F_j=e^{i\phi_j}$, $S_j=e^{\sigma_j}$, and $T_j=e^{-\tau_j}$, where
$j=1,\dots,n-5$ enumerates the internal tetragons. In
momentum-twistor space, these
transformations are again realized by the
unique $\grp{SL}(4)$ matrices $M\supup{int}_j$
satisfying
\begin{equation}
\bigbrk{Z_{j+4},Z_2,Z\supup{int}_{j+3},Z\supup{int}_{j+4}}.M\supup{int}_j
=
\sqrt{F_j}\diag\bigbrk{1/F_jS_j,S_j/F_j,T_j,1/T_j}.\bigbrk{Z_{j+4},Z_2,Z\supup{int}_{j+3},Z\supup{int}_{j+4}}
\label{eq:Mjcond}
\end{equation}
A family of conformally inequivalent null $n$-gons is now obtained by
successively acting with the stabilizing matrices $M\supup{int}_j$,
$j=n-5,\dots,1$ on the momentum twistors $Z_{j+5},\dots,Z_n,Z_1$ that
parametrize cusps below the $j$'th internal tetragon.

In \secref{sec:mrlrelations}, the $2\to(n-2)$ multi-Regge limit was
defined by $\varepsilon_j\to0$, with $\tilde u_j$ and
$\brk{1-u_{j-2,j+1}}^2/\varepsilon_j$ finite, for all $j=4,\dots,n-2$.
This is equivalent to
\begin{equation}
U_{i+2,n}\to0\,,
\qquad
U_{i+3,1}/U_{i+2,n}\text{ finite}\,,
\qquad
\brk{1-U_{i+1,i+4}}/U_{i+2,n}\text{ finite}\,.
\end{equation}
In the above tessellation parameters, this limit is attained by $T_i\to0$,
$S_i/T_i$ fixed, for all $i=1,\dots,n-5$. This can be understood in a similar
fashion as for the hexagon, see \figref{fig:tessellation}:
First, letting $T_{n-5}\to0$, $S_{n-5}/T_{n-5}$ fixed, takes
$x_n\to x_{n-1}$, which means $U_{j,n-1}\to1$ for $j=2,\dots,n-4$. At
the same time, it implies that $x_{n-2,n}$ and $x_{n-1,1}$ become
lightlike, which means $U_{n-3,n}$ and $U_{n-2,1}$ go to zero.
Subsequently letting $T_{n-6}\to0$, $S_{n-6}/T_{n-6}$ fixed, takes
$x_{n-1}\to x_{n-2}$, which implies $U_{j,n-2}\to1$ for
$j=2,\dots,n-5$, and $U_{n-4,n}\to0$, $U_{n-3,1}\to0$. This sequence
continues: Taking $T_i\to0$, $S_i/T_i$ fixed, implies $x_{i+5}\to
x_{i+4}$, and thus $U_{i+2,n}\to0$, $U_{i+3,1}\to0$, and
$U_{j,i+4}\to1$ (all at the same rate) for $j=2,\dots,i+1$.
Taking all $T_i\to0$ with all $S_i/T_i$ fixed completes the full
multi-Regge limit. We find the following astonishingly simple relation
between the tessellation parameters $F_i$, $S_i/T_i$ and the
multi-Regge kinematic variables $w_i$, $\bar w_i$ introduced in
\secref{sec:mrlrelations}:
\begin{equation}
F_i^2=b_i\frac{\bar w_{i+3}}{w_{i+3}}\,,
\qquad
\frac{S_i^2}{T_i^2}=\frac{c_i}{w_{i+3}\,\bar w_{i+3}}\,.
\label{eq:mrlparamid}
\end{equation}
Up to $\order{T_i^4}$, the large and small cross ratios expand to
\begin{equation}
U_{i+1,i+4}=1-\frac{(1+w_{i+3})(1+\bar w_{i+3})}{w_{i+3}\,\bar w_{i+3}}\,a_i\,T_i^2\,,
\quad
U_{i+2,n}=\frac{a_i\,T_i^2}{w_{i+3}\,\bar w_{i+3}}\,,
\quad
U_{i+3,1}=a_i\,T_i^2\,,
\label{eq:crmrlexp}
\end{equation}
As for the hexagon, the coefficients $a_i$, $b_i$, and $c_i$ are
numerical constants whose values depend on the choice of reference
$n$-gon. Experimentally, we find that one can always construct a
reference $n$-gon such that $b_i=1$, $c_i=-1$ for all $i=1,\dots,n-5$ in the above relations.

\paragraph{Explicit Construction.}

The parametrization explained above can be based on
arbitrary reference polygons (for example parametrized by $n$ random
momentum twistors). However, it is computationally very advantageous
to start with numerically simple reference polygons. In the following,
we outline how to construct such simple reference polygons.
Explicit parametrizations for
up to ten particles can be found in the
\mathematica\ file \texttt{mrlparam.m} attached to this publication.

Conformally inequivalent null hexagons can be parametrized as follows
(\textit{cf.}~\figref{fig:hexagon}):
\begin{align}
Z_1 &= \brk{1, 0, 4, 1}.M_1\,, &
Z_3 &= \brk{1, 0, 1, 1}\,, &
Z_5 &= \brk{0, 1, 0, 0}\,, &
Z\supup{int}_4 &= \brk{0, 0, 0, 1}\,,\nn\\
Z_2 &= \brk{1, 0, 0, 0}\,, &
Z_4 &= \brk{0, -1, 0, 1}\,, &
Z_6 &= \brk{0, 1, 1, 0}.M_1\,, &
Z\supup{int}_5 &= \brk{0, 0, 1, 0}\,,
\label{eq:hexaparam}
\end{align}
where $M_1$ stabilizes the internal tetragon
$\brc{Z_5,Z_2,Z\supup{int}_4,Z\supup{int}_5}$,
\begin{equation}
M_1(F,S,T)=\sqrt{F}\diag\bigbrk{S/F, 1/(F S), 1/T, T}\,.
\label{eq:M1}
\end{equation}
The reference hexagon (the above hexagon for $F=S=T=1$) can be
obtained from an arbitrary null hexagon by first picking a conformal
frame in which $\brc{Z_2,Z_5,Z\supup{int}_4,Z\supup{int}_5}$
take the above values, then applying a conformal transformation that
preserves the internal triangle and that takes $\brc{Z_3,Z_4}$ to the
above values, and finally choosing an origin for the
transformation $M_1$ such that $\brc{Z_6,Z_1}$ take the above values.
The $2\to4$ multi-Regge limit is attained by letting $T\to0$ with
$S/T$ fixed. Setting
\begin{equation}
F=\frac{\sqrt{\bar w_4}}{\sqrt{\vphantom{\bar w_4}w_4}}\,,
\qquad
\frac{S}{T}=-\frac{1}{\sqrt{\vphantom{\bar w_4}w_4}\sqrt{\bar w_4}}\,,
\end{equation}
the independent cross ratios up to $\order{T^4}$ expand to
\begin{equation}
U_{25}=1-\frac{(1+w_4)(1+\bar w_4)}{w_4\,\bar w_4}\,\frac{T^2}{4}\,,
\qquad
U_{36}=\frac{1}{w_4\,\bar w_4}\,\frac{T^2}{4}\,,
\qquad
U_{41}=\frac{T^2}{4}\,,
\end{equation}
that is,~\eqref{eq:mrlparamid,eq:crmrlexp} is satisfied with
$a_1=1/4$, $b_1=1$, $c_1=-1$.

A convenient parametrization of the null heptagon can be obtained by
extending the above null hexagon such that
the edge $x_{61}$ of the hexagon becomes the internal null line from
cusp $x_6$ to line $x_{12}$ of the heptagon. That is,
$Z_1$ of the hexagon becomes $Z_6\supup{int}$ of the heptagon.
To the configuration~\eqref{eq:hexaparam} one only needs to add a
new momentum twistor $Z_7$ that lies on the line
$(Z_6,Z\supup{int}_6)$, and a new momentum twistor $Z_1$ that lies in
the plane $(Z\supup{int}_6,Z_2,Z_3)$. This makes three new degrees of
freedom, which can be mapped to the parameters $F_2$, $S_2$,
$T_2$ of the new internal tetragon. A useful choice yields
\begin{align}
Z_1 &= \brk{3, 0, 12, 4}.M_2.M_1\,, &
Z_5 &= \brk{0, 1, 0, 0}\,, &
Z\supup{int}_4 &= \brk{0, 0, 0, 1}\,,\nn\\
Z_2 &= \brk{1, 0, 0, 0}\,, &
Z_6 &= \brk{0, 1, 1, 0}.M_1\,, &
Z\supup{int}_5 &= \brk{0, 0, 1, 0}\,,\nn\\
Z_3 &= \brk{1, 0, 1, 1}\,, &
Z_7 &= \brk{-1, 4, 0, -1}.M_2.M_1\,, &
Z\supup{int}_6 &= \brk{1, 0, 4, 1}.M_1\,,\nn\\
Z_4 &= \brk{0, -1, 0, 1}\,, &
& &
\label{eq:heptaparam}
\end{align}
where $M_1$ is the matrix~\eqref{eq:M1}, but with arguments $F_1$,
$S_1$, $T_1$, and
\begin{equation}
M_2(F_2,S_2,T_2)=
\sqrt{F_2}\begin{pmatrix}
S_2/F_2 & 0 & 0 & 0 \\
0 & 1/(F_2 S_2) & 1/(F_2 S_2) - T_2 & 0 \\
0 & 0 & T_2 & 0 \\
(F_2 - S_2 T_2)/(F_2 T_2) & 0 & 4(1 - T_2^2)/T_2 & 1/T_2
\end{pmatrix}
\label{eq:M2}
\end{equation}
stabilizes the new internal tetragon formed by
$\brc{Z_6,Z_2,Z\supup{int}_5,Z\supup{int}_6}\big|_{M_1=\mathrm{id}}$
according to~\eqref{eq:Mjcond}.
As described above, the $2\to5$ multi-Regge limit of the heptagon is obtained
by letting $T_1,T_2\to0$ with $S_1/T_1$, $S_2/T_2$ fixed. Identifying
\begin{equation}
F_i=\frac{\sqrt{\bar w_{i+3}}}{\sqrt{\vphantom{\bar w_{i+3}}w_{i+3}}}\,,
\qquad
\frac{S_i}{T_i}=-\frac{1}{\sqrt{\vphantom{\bar w_{i+3}}w_{i+3}}\sqrt{\bar w_{i+3}}}\,,
\qquad
i=1,2
\end{equation}
yields for the independent cross ratios up to $\order{T_i^4}$:
\begin{align}
U_{25}&=1-\frac{(1+w_4)(1+\bar w_4)}{w_4\,\bar w_4}\,\frac{T_1^2}{4}\,,
&
U_{37}&=\frac{1}{w_4\,\bar w_4}\,\frac{T_1^2}{4}\,,
&
U_{41}&=\frac{T_1^2}{4}\\
U_{36}&=1-\frac{(1+w_5)(1+\bar w_5)}{w_5\,\bar w_5}\,\frac{T_2^2}{4}\,,
&
U_{47}&=\frac{1}{w_5\,\bar w_5}\,\frac{T_2^2}{4}\,,
&
U_{51}&=\frac{T_2^2}{4}\,,
\end{align}
that is,~\eqref{eq:mrlparamid,eq:crmrlexp} again is satisfied with
$a_i=1/4$, $b_i=1$, $c_i=-1$.

Iterating this procedure, one can obtain convenient parametrizations of null
polygons with any number of edges. In the \mathematica\ file
\texttt{mrlparam.m}
attached to this publication, we provide explicit parametrizations for which the multi-Regge
limit~\eqref{eq:crmrlexp} is attained upon the
identification~\eqref{eq:mrlparamid} with $a_i=1/4$, $b_i=1$, $c_i=-1$, for
up to ten particles.

\bibliographystyle{nb}
\bibliography{multiregge}

\end{document}